\documentclass[prl,floatfix,superscriptaddress,twocolumn]{revtex4}
\usepackage{latexsym}
\usepackage{amstext,amsfonts}
\usepackage{epsfig}

\begin{document}

\title{Interdependent networks: Reducing the coupling strength leads to a change from a first to second order percolation transition}

\author{Roni Parshani}
\affiliation{Minerva Center \& Department of Physics, Bar-Ilan University, Ramat Gan, Israel}
\author{Sergey V. Buldyrev} \affiliation{Center for Polymer
 Studies and Dept. of Physics, Boston Univ., Boston, MA 02215 USA}
\affiliation{Department of Physics, Yeshiva University, 500 West 185th
 Street, New York, New York 10033, USA}
\author{Shlomo Havlin}
\affiliation{Minerva Center \& Department of Physics, Bar-Ilan University, Ramat Gan, Israel}

\date{\today}

\begin{abstract}
We study a system composed from two interdependent networks $A$ and $B$, 
where a fraction of the nodes in network $A$ depends on the nodes of network $B$ 
and a fraction of the nodes in network $B$ depends on the nodes of network $A$. 
Due to the coupling between the networks when nodes in one network fail they cause dependent nodes
in the other network to also fail. This invokes an iterative cascade
of failures in both networks. When a critical fraction of nodes fail
the iterative process results in a percolation phase transition that
completely fragments both networks. We show both analytically and numerically that reducing the
coupling between the networks leads to a change from a first order percolation phase transition to
a second order percolation transition at a critical point. The scaling of the
percolation order parameter near the critical point is characterized by the critical exponent $\beta=1$.

\end{abstract}
\maketitle

Most of the research on networks has concentrated on the limited case of a single 
network \cite{barasci,bararev,PastorXX,mendes,cohena} 
while real world systems are composed from many interdependent networks that interact 
with one another \cite{rinaldi,laprie,panzieri}. As a real example , 
consider a power-network and an Internet communication network that are coupled together. 
The Internet nodes depend on the power stations for electricity while
the power stations depend on the Internet for control \cite{rosato}.

\begin{figure}
\begin{center}
\epsfig{file=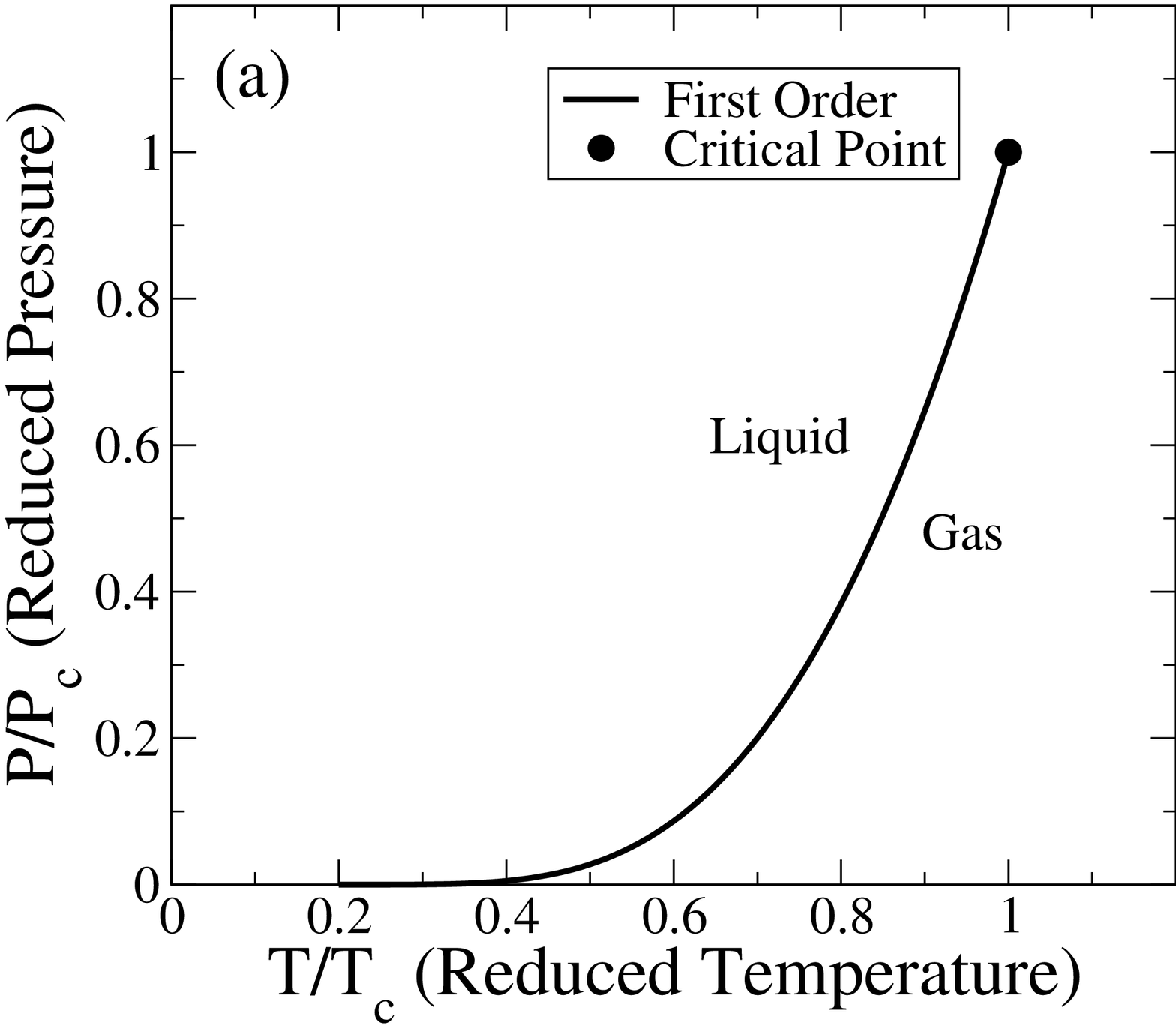,width=4cm}
\epsfig{file=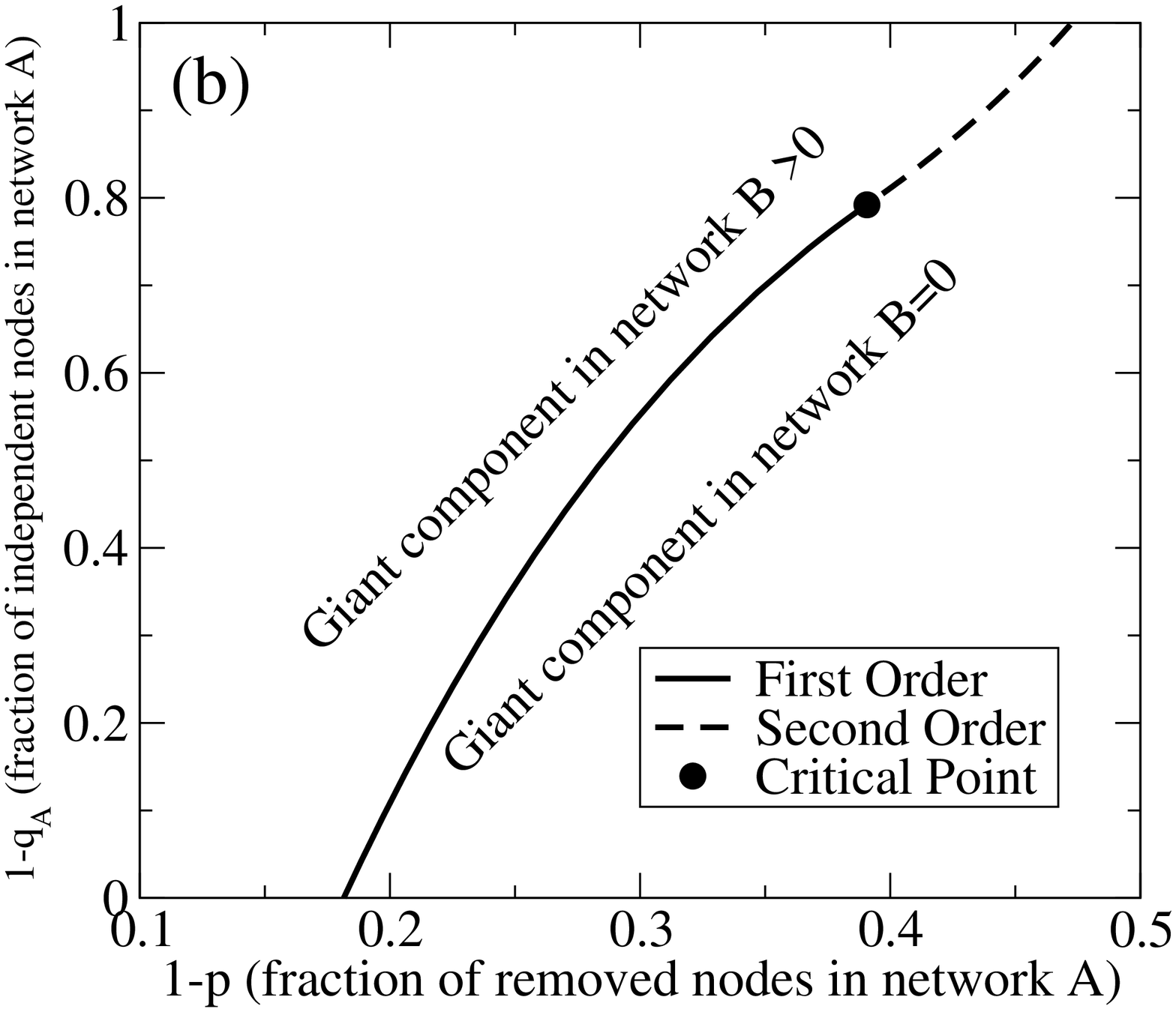,width=4cm}
\end{center}
\caption{(a) The van der Waals phase diagram. Along the liquid-gas 
equilibrium line the order parameter (density) abruptly
changes from a low value in the gas phase to a high value in the
liquid phase. At the critical point($P_c,T_c$)  the order parameter 
changes continuously as function of temperature if the pressure is kept constant at the 
critical value, but its derivative (compressibility) diverges. This is a
characteristic of the second order phase transition. (b) The 
percolation phase transition for two interdependent networks as
obtained from the numerical solution of system (\ref{systemf3}) for
$q_B=1$ and $a=b=3$. Here $1-p$, the fraction of removed nodes from
network A, plays the role of temperature. (As $1-p$ increases, the
disorder increases.)  The fraction $1-q_A$ of independent nodes in
network $A$ plays the role of pressure.  (As $1-q_A$ increases the
stability of network A increases.)  Below the critical point, the
system undergoes a first order phase transition at which, $\beta_\infty$, the fraction of nodes in the giant component 
of network B abruptly changes from a finite value to zero. 
As we approach the critical point, $\beta_\infty \to 0$. 
Above the critical point, the system undergoes a second order transition where the giant component continuously approaches zero.}
\label{first_second_transition}
\end{figure}

We show that introducing interactions between networks is analogous to
introducing interactions among molecules in the ideal gas model. Interactions among molecules 
lead to the replacement of the ideal gas law by the Van der Waals equation
that predicts a liquid-gas first order phase transitions line ending at a
critical point characterized by a second order transition (Fig.\ref{first_second_transition}(a)). 
Similarly, interactions between networks give rise to a first order percolation phase transition line that changes to a second order
transition, as the coupling strength between the networks is reduced (Fig.\ref{first_second_transition}(b)).   
At the critical point the first order line merges with the second order line, near which the order parameter (the size of giant component)
scales linearly with the distance to the critical point, leading to the critical exponent $\beta=1$.

In interdependent networks, nodes from one network depend on nodes from
another network. Consequently, when nodes from one network fail they
cause nodes from another network to also fail. If the connections
within each network are different, this may trigger a recursive
process of a cascade of failures that can completely fragments both
networks.  Recently, Buldyrev et al \cite{Buldyrev} studied the coupling between
two $N$ node networks $A$ and $B$ assuming the following restrictions: (i) Each and every node in network $A$
depends on {\it one} node from network $B$ and vice versa. (ii) If node $A_i$ depends on node $B_i$ 
then node $B_i$ depends on node $A_i$. They show that for such a model when
a critical fraction of the nodes in one network fail, the system
undergoes a first order phase transition due to the recursive
process of cascading failures.

However, when examining the features of real interdependent networks such as the 
power network and the communication network presented above, 
we observe that in practice not all nodes of
network $A$ depend on network $B$ and vice versa. We therefore
introduce a general model that is applicable to many real
networks.  The model consists of two networks A and B with the number of
nodes $N_A$ and $N_B$, respectively. Within network $A$, the nodes are
randomly connected by A-edges with degree distribution $P_A(k)$, while
the nodes in network B are randomly connected by B-edges with degree
distribution $P_B(k)$.  In our model a fraction $q_A$ of network
$A$ nodes depends on the nodes in network $B$ and a fraction $q_B$ of
network $B$ nodes depends on the nodes in network $A$. We find that for 
strong coupling (large values of $q_A$ and $q_B$) the networks
undergo a first order transition while for a weak coupling they
undergo a second order phase transition.  Even for the case of weak
coupling in which a second order percolation transition occurs, the
system still disintegrates in an iterative process of cascading failures
unlike a regular second order percolation transition for a single
network.

\begin{figure} [t]
\begin{center}
\epsfig{file=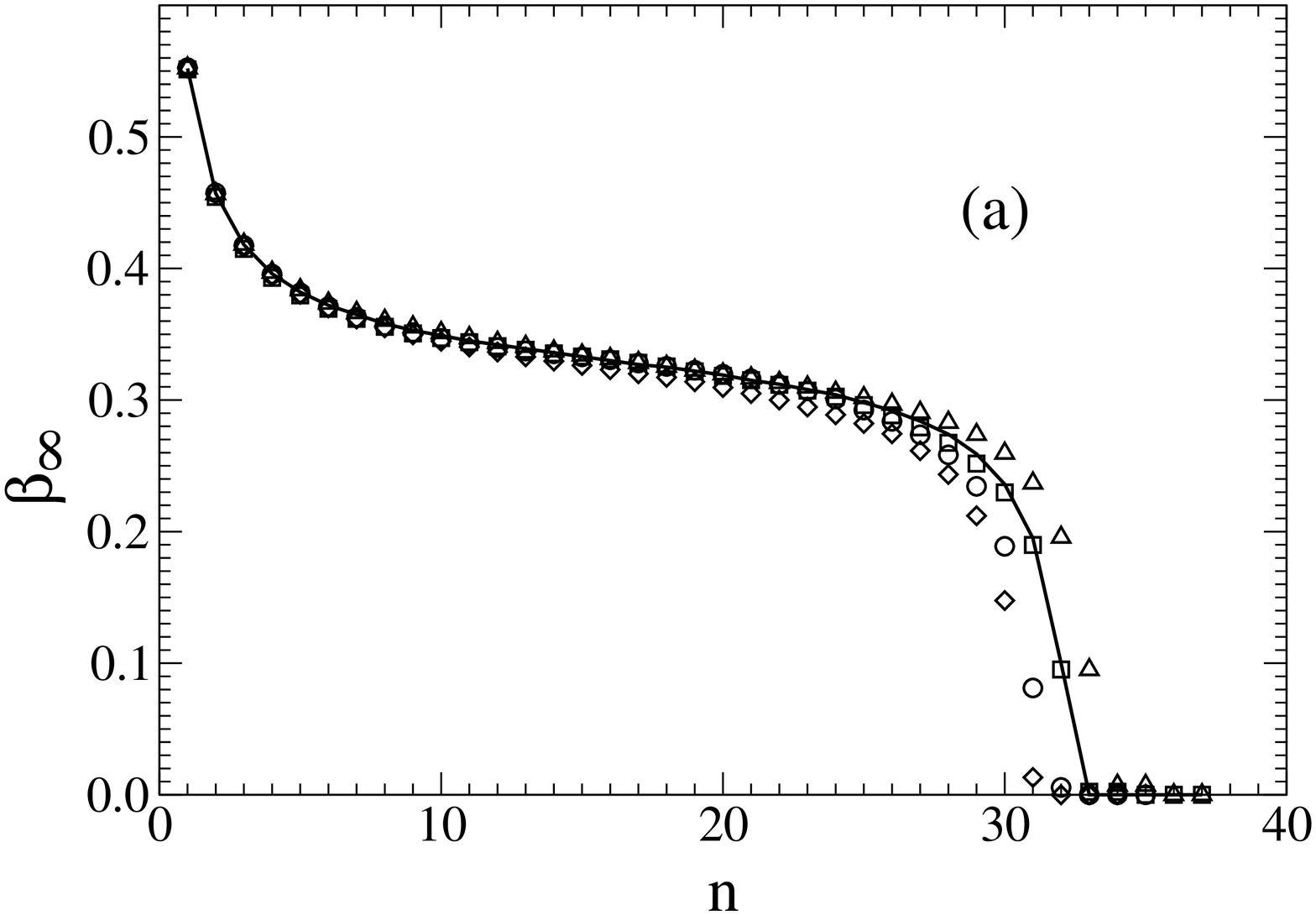,height=4cm,width=4cm}
\epsfig{file=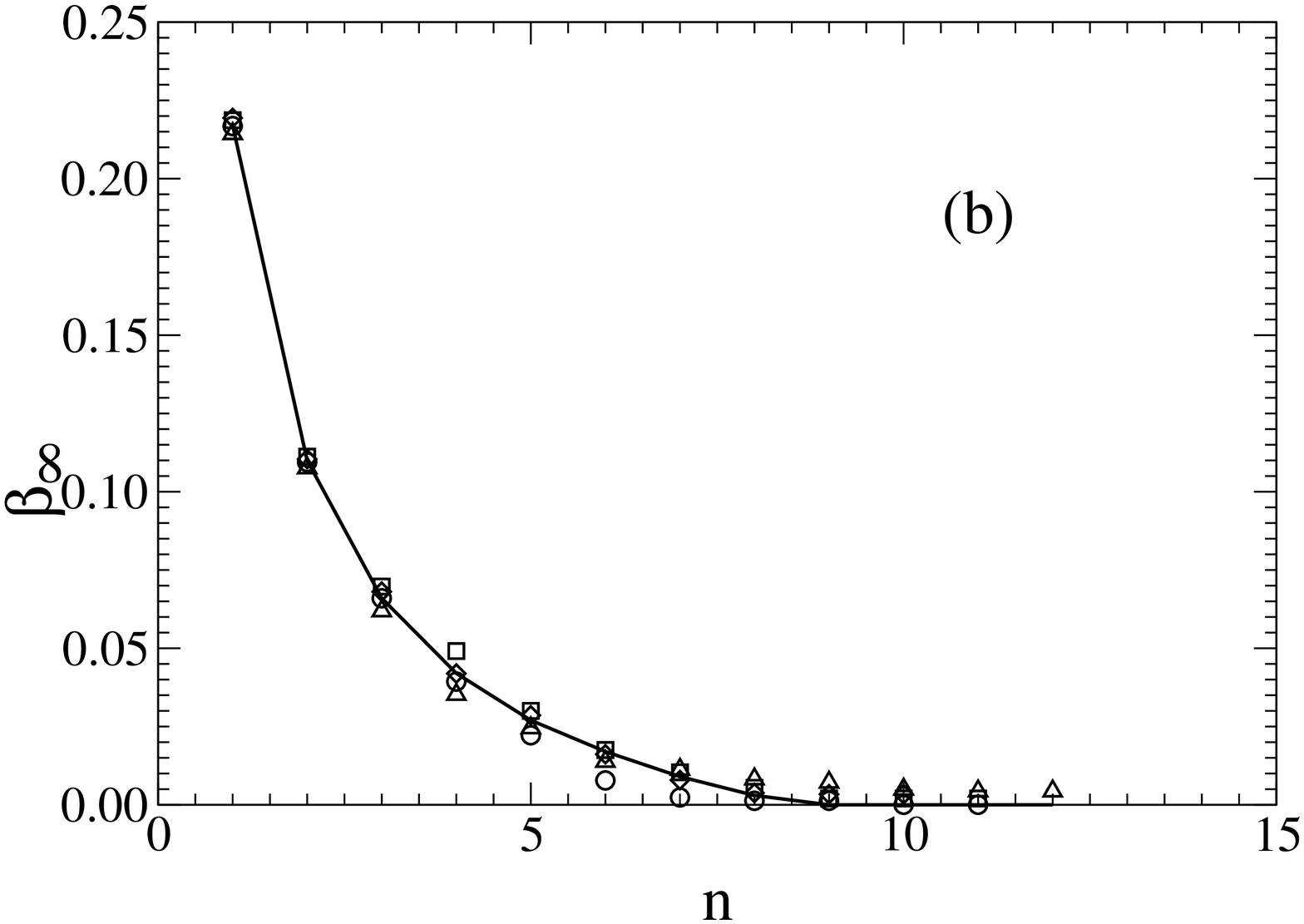,height=4cm,width=4cm}
\epsfig{file=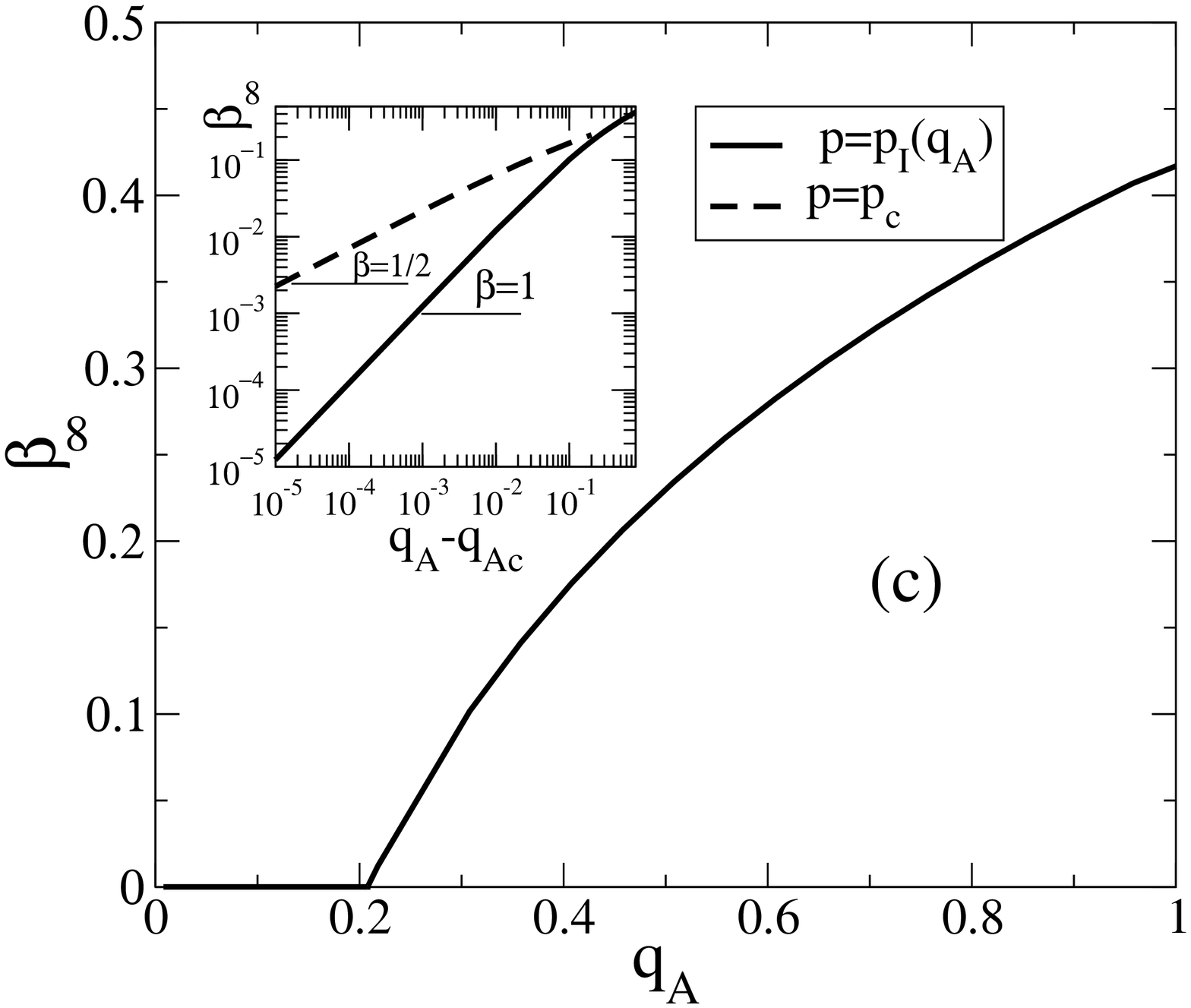,height=4cm,width=4cm}
\epsfig{file=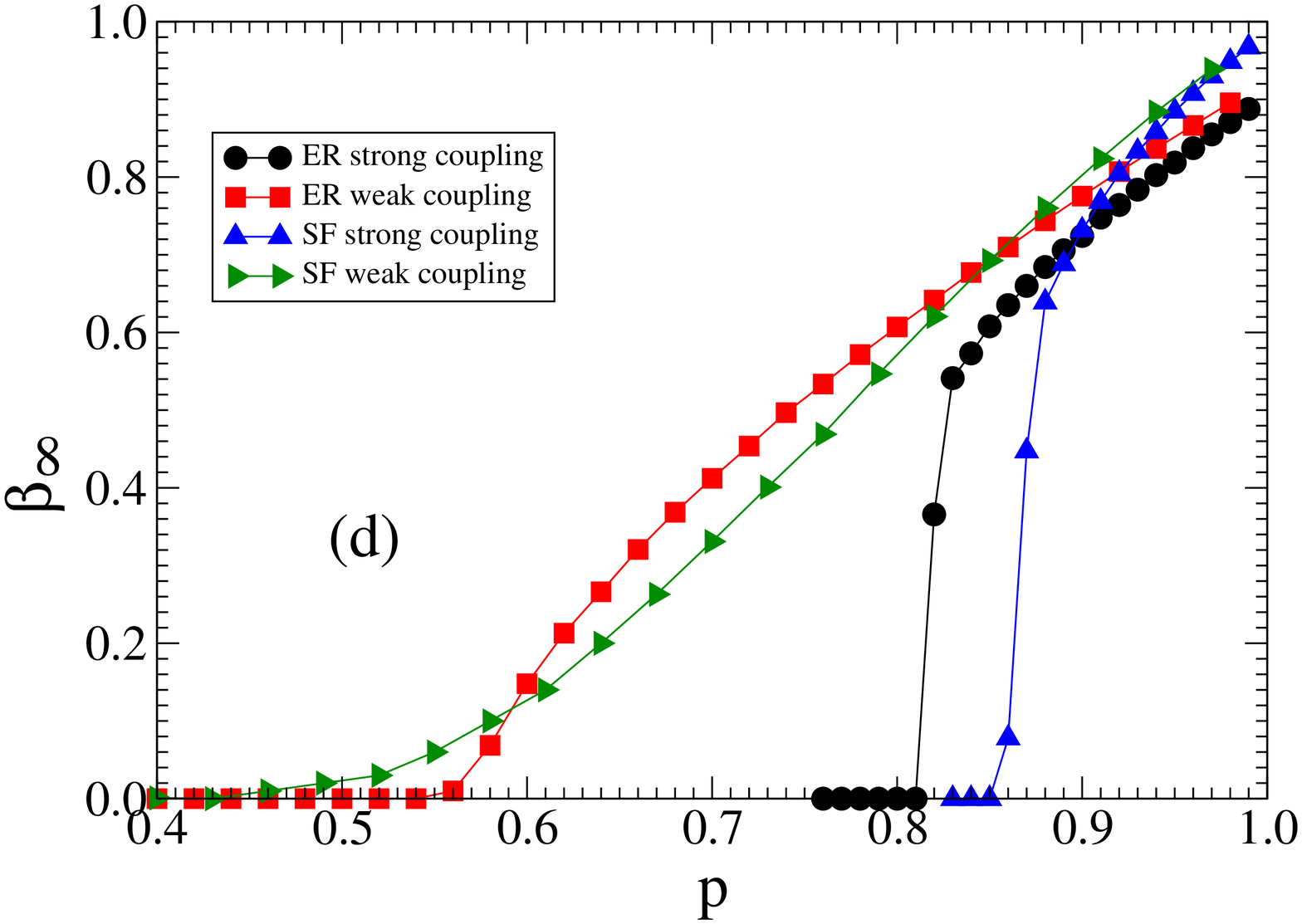,height=4cm,width=4cm}
\end{center}
\caption{An iterative process of failures for ER networks of size
$N_A=N_B=8\times10^5$ (a) A first order iterative process for 
$p=0.7455$, $a=b=2.5$, $q_A=0.7$ and $q_B=0.6$.  (b) A second order iterative process for 
$p=0.605$, $a=b=2.5$, $q_A=0.2$ and $q_B=0.75$. Symbols
represent simulation results for different random realizations of the
networks. Solid lines represent the solution of system (\ref{systemf0}).
(c) The fraction of nodes in network's $B$ giant component, $\beta_\infty$, as function of
$q_A$ computed at $p=p_I(q_A)$, the line of the first order phase transition. 
The results are obtained by solving system
(\ref{systemf3}) with additional condition $df_A/df_B \times
df_B/df_A=1$ for $a=3, b=3, q_B=1$. Inset: The same results (solid
line) as function of $|q_A-q_{A_c}|$  yield a straight line with
slope $\beta=1$ in double logarithmic scale. 
If $q_A$ is changed but $p=p_c$ is kept constant we obtain a straight line with slope $\beta=0.5$ (dashed line).  
(d) Simulation results for the phase transition of
$\beta_{\infty}$ as a function of $p$ for $N=50K$.  For strong coupling
between the networks we observe a jump in $\beta_{\infty}$ as expected
in the first order phase transition (ER(circle) and SF(up rectangle)).
For weak coupling between the networks the change in $\beta_{\infty}$
is gradual as expected for the second and higher order phase
transitions (ER(square) and SF(right rectangle)).}
\label{iterative_process}
\end{figure}
%
%
The iterative process of cascading failures starts with randomly
removing a fraction $1-p$ of network A nodes and all the A-edges that
are connected to them.  Due to the interdependence between the networks,
the nodes in network B that depend on removed A-nodes are also removed
together with the B-edges that are connected to them.  As nodes and
edges are removed, each network breaks up into connected components,
that we call clusters. We assume that when
the network is fragmented, the nodes belonging to the giant component
connecting a finite fraction of the network, are still functional,
while nodes that are parts of the remaining small clusters become
non-functional.  Since each network is connected differently, the
nodes that become non-functional on each step are different for both
networks. This leads to the removal of more dependent nodes from the
coupled network and so on.

Next we present the formalism for the cascade process step by step. We define $p_A$ and $p_B$ as the fraction of nodes belonging to the giant components of network $A$ and $B$ respectively.
The remaining fraction of network $A$ nodes after an initial removal of $1-p$ is $\alpha'_1\equiv p$.
The initial removal of nodes will disconnect additional nodes from the giant cluster.
The remaining functional part of network $A$ therefore contains a fraction $\alpha_1=\alpha'_1p_A(\alpha'_1)$ of the network nodes. Since a fraction $q_B$ of nodes from network $B$ 
depend on nodes from network $A$, the number of nodes in network $B$ that become non functional is $(1-\alpha_1)q_B = q_B(1-\alpha'_1p_A(\alpha'_1))$.
Accordingly, the remaining fraction of network $B$ is $\beta'_1=1-q_B(1-\alpha'_1p_A(\alpha'_1))$ 
and the fraction of nodes in the giant component of network $B$  is $\beta_1=\beta'_1p_B(\beta'_1)$.

Following this approach we can construct the sequence, $\alpha_n$ and
$\beta_n$, of giant components, and the sequence, $\alpha'_n$ and
$\beta'_n$, of the remaining fraction of nodes at each stage of the
cascade of failures.  The general form is given by: \\
$\alpha'_1\equiv p$, $\alpha_1=\alpha'_1 p_A(\alpha'_1)$,\\
$\beta'_1=1-q_B(1-p_A(\alpha'_1)p)$, $\beta_1=\beta'_1 p_B(\beta'_1)$,\\ 
$\alpha'_2=1 - \alpha'_1[1-q_A(1-p_B(\beta'_1))]$, $\alpha_2=\alpha'_2 p_A(\alpha'_2) \ldots$\\
$\alpha'_{m}=p[1-q_A(1-p_B(\beta'_{m}))]$, $\alpha_{m}=\alpha'_{m} p_A(\alpha'_{m})$,\\ 
$\beta'_{m}=1-q_B(1-p_A(\alpha'_{m})p)$, $\beta_{m}=\beta'_{m} p_B(\beta'_{m})$.\\

\noindent To determine the state of the system at the end of the
cascade process we look at $\beta'_{m}$ and $\alpha'_{m}$ at the limit
of $m \to \infty$.  This limit must satisfy the equation
$\alpha'_{m}$=$\alpha'_{m+1}$ (or $\beta'_{m}$=$\beta'_{m+1}$) since
eventually the clusters stop fragmenting and the fractions of
randomly removed nodes at step $m$ and $m+1$ are equal.  Denoting
$\beta'_{m}=y$ and $\alpha'_{m}=x$ we arrive to a system of two
equations with two unknowns:
\begin{equation}
\left\{
\begin{array}{lr}
y=1-q_B(1-p_A(x)p)\\
x=p[1-q_A(1-p_B(y))].
\end{array}
\right.\
\label{system}
\end{equation}
The model can be solved analytically
using the apparatus of generating functions.  The generating functions
will be defined for network A while similar equations describe network
B.  As in Refs.~\cite{Newman,Shao} we will introduce the generating
function of the degree distributions $G_{A0}(\xi)=\sum_k P_A(k) \xi^k$.
Analogously we will introduce the generating function of the underlining
branching processes, $G_{A1}(\xi)=G'_{A0}(\xi)/ G'_{A0}(1)$.
Random removal of fraction $1-p$ of nodes will change the degree
distribution of the remaining nodes, so the generating function of the
new distribution is equal to the generating function of the original
distribution with the argument equal to $1-p(1-\xi)$
\cite{Newman}. The fraction of nodes that belong to the giant
component after the removal of $1-p$ nodes is \cite{Shao}:
\begin{equation}
p_A(p)=1-G_{A0}[1-p(1-f_A)],
\label{e:p_A}
\end{equation}
where $f_A=f_A(p)$ satisfies a transcendental equation
\begin{equation}
f_A=G_{A1}[1-p(1-f_A)].
\label{e:f_A}
\end{equation}
In case of two ER networks, whose degrees are Poisson-distributed
\cite{er1,bollo}, the problem can be solved explicitly. Suppose that
the average degree of the network A is $a$ and the average degree of
the network B is $b$.  Then, $G_{A1}(\xi)=G_{A0}=\exp[a(\xi-1)]$ and
$G_{B1}(\xi)=G_{B0}=\exp[b(\xi-1)]$. Accordingly, $p_B(x)=1-f_B$ and $p_A(x)=1-f_A$  
and therefore system (1) becomes
\begin{equation}
\left\{
\begin{array}{lr}
x=p[1-q_Af_B]\\
y=1-q_B(1-p[1-f_A]),
\end{array}
\right.\
\label{systemf0}
\end{equation}

where $f_A$ and $f_B$ satisfy the
transcendental equations 
\begin{equation}
\left\{
\begin{array}{lr}
f_A=\exp[ax(f_A-1)]\\
f_B=\exp[by(f_B-1)].
\end{array}
\right.\
\label{systemf1}
\end{equation}

The fraction of nodes in the giant components of networks A and B respectively, at the end of the cascade process are given by $\alpha_\infty=p(1-f_A)(1-q_A f_B)$ and  $\beta_\infty=(1-f_B)(1+q_B(1-p)-p q_B f_A)$. 
Fig.(\ref{iterative_process}) shows excellent agreement between computer simulations of the cascade failures  
and the numerical results obtained by solving systems (\ref{systemf0}) and (\ref{systemf1}). 
Excluding $x$ and $y$ from systems (\ref{systemf0}) and (\ref{systemf1}), 
we obtain a system:
\begin{equation}
\left\{
\begin{array}{lr}
f_A=e^{-ap(f_A-1)(q_A f_B-1)}\\
f_B=e^{-b(q_B(1-p[1-f_A])-1)(f_B-1)}.
\end{array}
\right.\
\label{systemf2}
\end{equation}
The first equation can be solved with respect to $f_B$ and the second equation can be solved with respect to $f_A$
\begin{equation}
\left\{
\begin{array}{lr}
f_B=\frac{1}{q_A}[1-\frac{\log{f_A}}{ap(f_A-1)}],f_A\neq 1;~\forall f_B,f_A=1\\
f_A=\frac{1}{q_B}[\frac{1+q_B(p-1)}{p}-\frac{\log{f_B}}{bp(f_B-1)}],f_B\neq 1;~\forall f_A,f_B=1
\end{array}
\right.\
\label{systemf3}
\end{equation}

\begin{figure}
\begin{center}
\epsfig{file=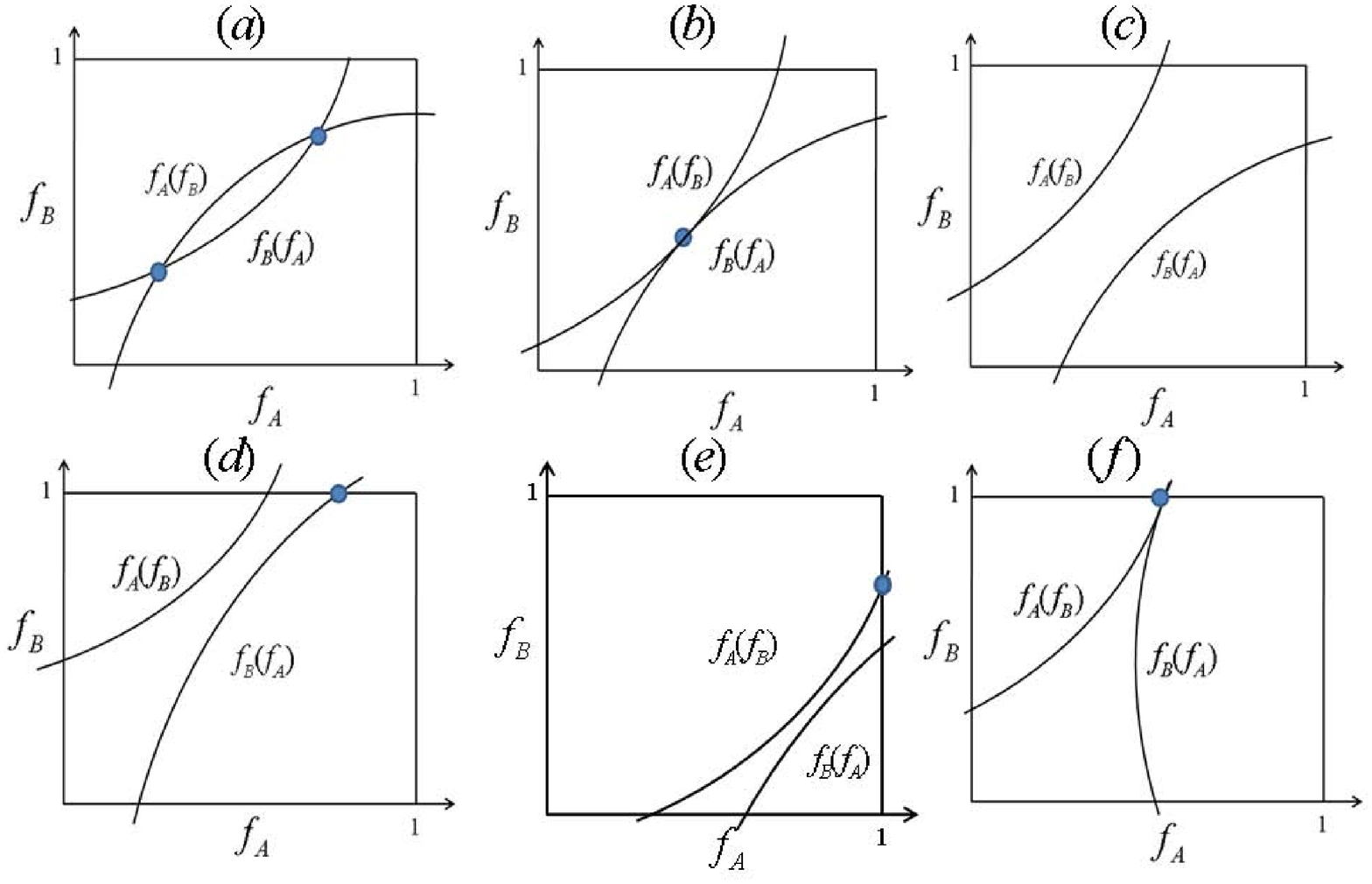,height=6.2cm,width=6.2cm}
\end{center}
\caption{Illustrations of the different graphical solutions of system (\ref{systemf3}) (see manuscript for detailed explanation of the different plots).}
\label{graphic_solution}
\end{figure}
The solutions of system (\ref{systemf3}) can be graphically presented
on a $f_A,f_B$ plane (Fig.~\ref{graphic_solution}).  The solutions are
presented as a crossing of either $f_B(f_A)$ or $f_A=1$ with
$f_B(f_A)$ or $f_A=1$ and are restricted to the square $0\leq f_A \leq
1$ ; $0\leq f_B \leq 1$.  There are three different possible
solutions: {\it (i)} The solution where the giant components of both
networks are zero ($f_A$=1 and $f_B$=1) as in
Fig.\ref{graphic_solution}(c). {\it (ii)} A solution for which only one
of the giant components of either network $A$ or $B$ is zero ($f_A=1$
and $f_B\not=1$ or $f_A\not=1$ and $f_B=1$) as in
Fig.\ref{graphic_solution}(d) (or Fig.\ref{graphic_solution}(e)).
{\it (iii)} A solution for which both networks have a non-zero giant
component ($f_A\not=1$ and $f_B\not=1$). This solution is given by the
lowest intersection point of the curves in
Fig.\ref{graphic_solution}(a). This solution may disappear in two
different scenarios. 

The first scenario is presented in
Fig.\ref{graphic_solution}(b) in which an infinitesimal change
$\bigtriangleup \vec{z} $ in the vector of the system parameters
$\vec{z}=(a,b,q_A,q_B,p)$ may lead to a first order phase transition
in which the size of one or both of the giant components changes discontinualy from a
finite value to zero: (Fig.\ref{graphic_solution}(a) $\rightarrow$
Fig.\ref{graphic_solution}(b) $\rightarrow$
Fig.\ref{graphic_solution}(c) or Fig.\ref{graphic_solution}(d), or
Fig.\ref{graphic_solution}(e)).  The condition for the first order
phase transition is $\frac{df_B(f_A)}{df_A}\frac{df_A(f_B)}{df_B}=1$
corresponds to the touching point of the two curves as
in Fig.\ref{graphic_solution}(b).  When adding this condition to the
two equations in system (\ref{systemf3}) we can find the three
unknowns $f_A=f_{A_I}$,$f_B=f_{B_I}$ and $p=p_I$ for given
$a,b,q_A,q_B$. Fixing $a,b,q_B$ will define a first order phase transition line 
$p=p_I(q_A)$ as function of $q_A$ [Fig.~\ref{first_second_transition}(b)].

The second scenario is presented in Fig.\ref{graphic_solution}(f). 
In this case (corresponding to $f_A<1, f_B=1$ or equivalently to $q_B>1-1/b$), 
$\beta_\infty$, continually decreases to zero, 
while $\alpha_\infty$ stays finite. 
This situation corresponds to the second order phase
transition that can be found by substituting $f_B=1$ into system (\ref{systemf3}). 
These two equations allow one to find $f_A=f_{A_{II}}$, and $p=p_{II}$ which for
fixed $a,b,q_B$ define a line of second order phase transitions
$p=p_{II}(q_A)$ as a function of $q_A$ [Fig.~\ref{first_second_transition}(b)]. 
   
\begin{figure}
\begin{center}
\epsfig{file=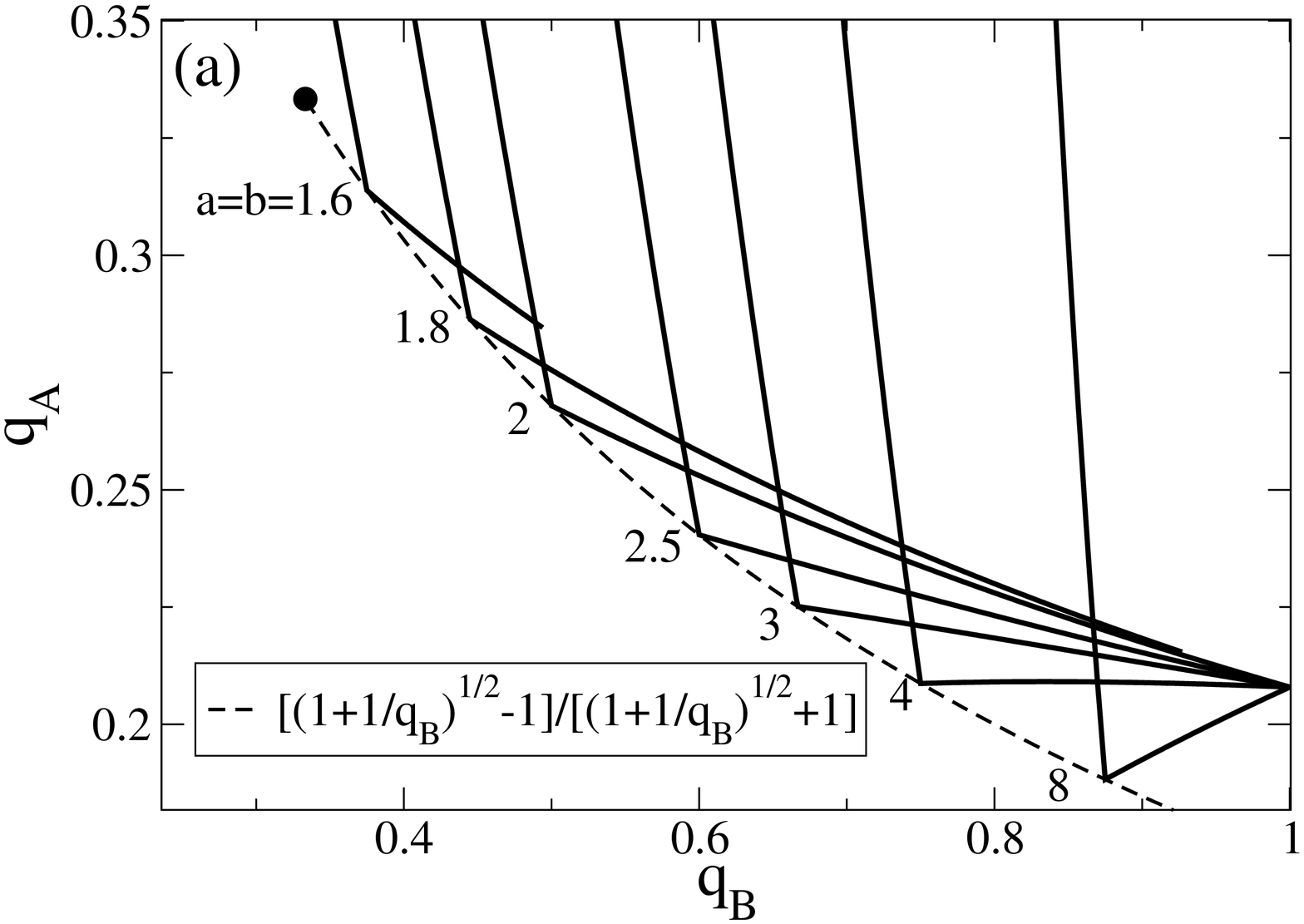,height=4cm,width=4cm}
\epsfig{file=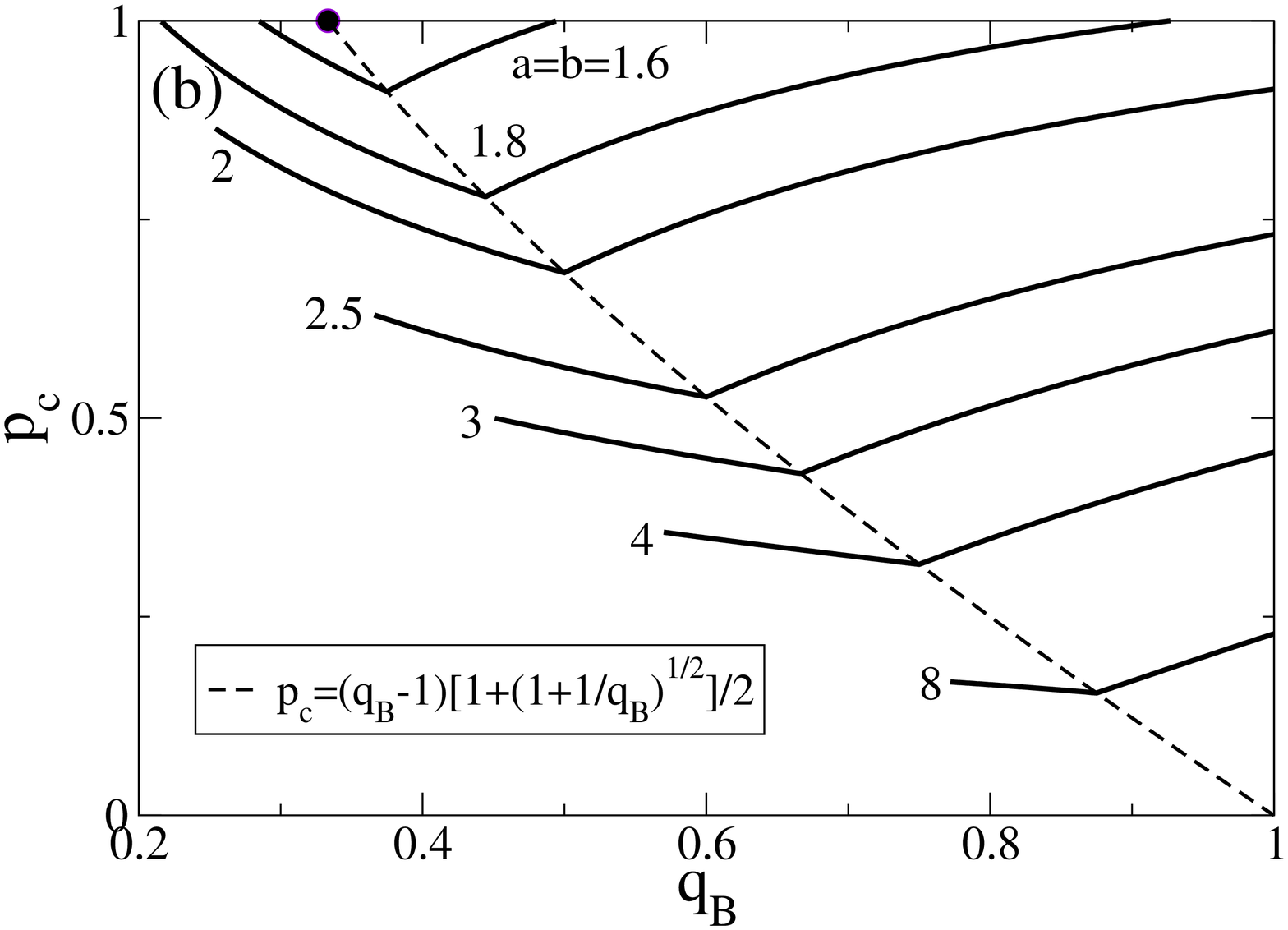,height=4cm,width=4cm}
\end{center}
\caption{The critical point parameters, $p_c$ and $q_{A_c}$, 
as functions of the coupling strength $q_B$ for ER networks are plotted for different values of $a=b$. 
(a) For $q_B=1$ the networks (with large degrees) have the same
critical coupling strength $q_{A_c}=\mathcal{W}[\exp(3)]=0.20794$. The
networks with small degrees do not have first order phase transitions
(no critical points for $q_B$), because for large values of
$q_B$, $p_c(q_B)>1$, which is unphysical.  The range of $q_B$ values
for which the first order phase transition exists shrinks as $a=b$
decreases and eventually disappears for $a=b=3/2$, when the
critical point exists only for $q_{A_c}=q_B=1/3$ and $p_c=1$. This point
is marked by a solid circle.}
\label{giant_cluster_transition}
\end{figure}

The line of the first order phase transitions merges with the line of
the second order phase transitions in a critical point which can be
found by adding to system (\ref{systemf3}) both the first order condition
$\frac{df_B(f_A)}{df_A}\frac{df_A(f_B)}{df_B}=1$ and the second order
condition $f_B=1$ or $f_A=1$. These four equations allow us to find the 
critical parameters $f_B=f_{B_c}$ or $f_A=f_{A_c}$, $p=p_c$ and
$q_A=q_{A_c}$ as functions of $a,b,q_B$. 
Fig.~\ref{giant_cluster_transition} presents the solution for $p_c(q_B)$ and
$q_{A_c}(q_B)$ for different values of $a$($=b$). 
The kink in the solutions occurs when both curves
tangentially intersect at $f_A=1,f_B=1$ which corresponds to $\tilde{q_B}=1-1/b$.
The minimal value of $p_c$ occurs exactly at the kink, defining the  
condition for the first order phase transition as $p_c(\tilde{q_B})<1$.
Thus the first order transition can exist only in dense networks with
sufficiently high average degrees, such that $4(a-1)(b-1)>1$. Low degree networks must
disintegrate in the second order phase transitions.

At the critical point the system can be reduced
to a single transcendental Lambert equation. For the most simple case $a/b=q_B=1$, we find that $f_{A_c}=1/z$,
$q_{A_c}=z-2$, $p_c=z/[a(z-1)]$ and $\alpha_\infty= (3-z)/a$, where
$z=\mathcal{W}[exp(3)]=2.20794$ satisfies the Lambert equation
$z\exp(z)=\exp(3)$. 

To find the critical exponent $\beta$ near the critical point
we express the order parameter $\beta_\infty(q_A)$ as function of
$q_A>q_{A_c}$ along the transition line $p=p_I(q_A)$
(inset of Fig.~\ref{iterative_process}(c)). Expanding $f_B$ in series of
$x=q_A-q_{A_c}$ we find that $\lim_{x\to 0}(1-f_B)/x =C>0$, indicating
that $\beta=1$. Interestingly, if one keeps $p=p_c$ constant and
changes only $q_A$, then $\lim_{x\to 0}(1-f_B)/\sqrt{x}=C'<0$
corresponding to $\beta=1/2$. The inset of Fig.~\ref{iterative_process}(c) 
confirms our analytical predictions numerically.

Although our analytical theory is developed for ER networks, the same
qualitative conclusions hold for randomly connected networks with
arbitrary degree distributions, since functions $p_A(x)$ and $p_B(y)$
can be expressed in terms of generating functions of these
distributions. Hence an analysis similar to Fig. 3 holds for any degree distributions. 
Computer simulations of interacting SF networks and ER networks presented in Fig.~\ref{iterative_process}(d)
support this analysis.

\noindent We thanks the European EPIWORK project, the Israel Science Foundation and the DTRA for financial support.
S.V.B. thanks the Office of the Academic Affairs of Yeshiva University for funding the Yeshiva
University high-performance computer cluster and acknowledges the
partial support of this research through the Dr. Bernard W. Gamson
Computational Science Center at Yeshiva College.

\end{document}